\documentclass[prl,twocolumn,showpacs,amssymb]{revtex4}
\usepackage{amsmath,bm}
\usepackage{graphicx}
\begin{document}

\title{Theory of Electromotive Force Induced by Domain Wall Motion}

\author{Shengyuan A. Yang}

\affiliation{Department of Physics, The University
of Texas, Austin, Texas, 78712-0264, USA}

\author{Di Xiao}
\affiliation{Department of Physics, The University of Texas, Austin,
Texas, 78712-0264, USA}

\author{Qian Niu}
\affiliation{Department of Physics, The University of Texas, Austin,
Texas, 78712-0264, USA}

\date{\today}

\begin{abstract}

We formulate a theory on the dynamics of conduction electrons in the
presence of moving magnetic textures in ferromagnetic materials. We
show that the variation of local magnetization in both space and
time gives rise to topological fields, which induce electromotive
forces on the electrons. Universal results are obtained for the emf
induced by both transverse and vortex domain walls traveling in a
magnetic film strip, and their measurement may provide clear
characterization on the motion of such walls.

\end{abstract}

\maketitle

The interplay between electron transport and magnetic dynamics is a
central problem of spintronics research.  It has been known that the
presence of a domain wall (DW) can change the electrical resistance
of a ferromagnetic conductor~\cite{greg,gork,tafu,duga}. It has also
been demonstrated that an electric current can drive a DW through
coupling between the conduction electrons and the local magnetic
moments~\cite{berg,tk,lizh,thia,barn1,ohe,bena,yama,klau,beac1,haya}.
The reverse of this effect, i.e., electron transport induced by a
moving DW, has been proposed by Berger \cite{berg1} based on
phenomenological arguments about twenty years ago. Recently there is
renewed interest in this effect and the result for 1D transverse
domain wall has been derived rigorously using various approaches
\cite{barn,duin,sasl}, connecting to Berry phase effects on electron
spins in a magnetic texture \cite{berr,brau,baza}. However,
realistic DW profiles and their motion are far more complicated than
the simple 1D model \cite{mcmi,klau1}, typically involving vortices.
So a general microscopic theory applicable to higher dimensions is
very much desired.

In this Letter, we provide such a theory within the framework of
semiclassical dynamics of electrons in a magnetic background which
varies slowly in both space and time. Indeed, the width of a typical
DW in a ferromagnetic nanowire is about a few hundred nanometers,
which is much larger than the electron Fermi wavelength, and the DW
speed is much smaller than the electron speed. One can therefore
consider the semiclassical formalism with adiabatic approximation
where the spin of the conduction electron follows the direction of
local spin vector. We find that Berry phase terms \cite{berr} enter
into the equation of motion as a pair of topological fields, acting
locally like electric and magnetic fields on the electrons.

While our theory naturally reproduces previous results on 1D
transverse walls, we consider carefully the effect of a moving
vortex wall. We predict both longitudinal and transverse voltage
with universal results which may be used to provide clear
characterization of the wall motion. Extra voltage due to
nonadiabatic effects is also estimated at the end of the paper, and
is found to be subdominant.

To construct our theory, we consider a ferromagnetic thin film,
which is taken to be the x-y plane. The time evolution of local
magnetization can be driven by a uniform applied magnetic field
(x-direction).  Our model Hamiltonian then takes the following form,
\begin{equation}
H=H_0[\bm{q}+(e/\hbar)\bm{A}(\bm{r})]-
J\hat{\bm{n}}(\bm{r},t)\cdot\bm{\sigma}-h\sigma_x.
\end{equation}
The first term is the bare Hamiltonian for a conduction electron,
$\bm{q}$ is the Bloch wave-vector, and $\bm{A}(\bm{r})$ is the
vector potential of the external magnetic field. The second term is
the \emph{s-d} coupling between a conduction electron and the local
d-electron spin along direction $\hat{\bm{n}}(\bm{r},t)$, and $J$ is
the s-d coupling strength. The last term represents the Zeeman
coupling between electron and the external magnetic field, with
$h=\frac{1}{2}g_s\mu_B B$.

To apply the semiclassical wave-packet formalism for the conduction
electrons \cite{sund}, we first write down the local Hamiltonian at
the center position of the electron wave-packet $\bm{r}_c$,
\begin{equation}
H_c=H_0[\bm{q}+(e/\hbar)\bm{A}(\bm{r}_c)]-K(\bm{r}_c,t)
\hat{\bm{n}}'(\bm{r}_c,t)\cdot\bm{\sigma},
\end{equation}
where $\hat{\bm{n}}'(\bm{r}_c,t)$ is the unit vector of the exchange
plus Zeeman field, while $K(\bm{r}_c,t)$ is its strength.  As
discussed by Sundaram and Niu \cite{sund}, we introduce the gauge
invariant crystal momentum $\bm{k}=\bm{q}+(e/\hbar)\bm{A}(\bm{r})$.
For now, we only consider the majority carriers whose spins are
polarized along $\hat{\bm{n}}'(\bm{r}_c,t)$, and spin minority
carriers are considered in the end. The position and time dependence
of the spinor wave function gives rise to Berry curvatures in space
and time, which can affect the dynamics of an electron wavepacket.
More specifically, we find that the equations of motion for the
wavepacket center are (subscript $c$ is dropped here)
\begin{equation}\label{xdot}
\dot{\bm{r}}=\frac{\partial \mathcal{E}_0}{\hbar\partial \bm{k}},
\end{equation}
\begin{equation}\label{kdot}
\dot{\bm{k}}=\frac{\partial K}{\hbar\partial
\bm{r}}-\frac{e}{\hbar}\dot{\bm{r}}\times\bm{B}-\dot{\bm{r}}\times
\bm{C}-\bm{D},
\end{equation}
where $\mathcal{E}_0$ is the Bloch band energy obtained from $H_0$.
Originated from Berry curvatures in real space and time, two new
fields $\bm{C}$ and $\bm{D}$ appear in the equations of motion. They
are entirely due to the spatial and temporal variation of local spin
textures. In terms of the spherical angles $(\theta,\phi)$
specifying the direction of $\hat{\bm{n}}'$, the fields $\bm{C}$ and
$\bm{D}$ are given by
\begin{equation}
\bm{C}(\bm{r},t)\equiv\frac{1}{2}\sin\theta\left(\nabla\theta\times\nabla\phi\right),
\end{equation}
\begin{equation}
\bm{D}(\bm{r},t)\equiv\frac{1}{2}\sin\theta\left(\frac{\partial
\phi}{\partial t}\nabla\theta-\frac{\partial \theta}{\partial
t}\nabla\phi\right).
\end{equation}
Field $\bm{C}$ is similar to the gyrovector used in the discussion
of Bloch line dynamics \cite{malo}. From their appearance in
equation (\ref{kdot}), we observe that $\bm{C}$ acts like a magnetic
field while $\bm{D}$ behaves like an electric field.  We note that
our theory applies locally to the dynamics of electrons, while
recent work of Ref.\cite{barn,duin} focuses more on the global
aspect of Berry phase effects.

\begin{figure}
\includegraphics[width=0.8\columnwidth]{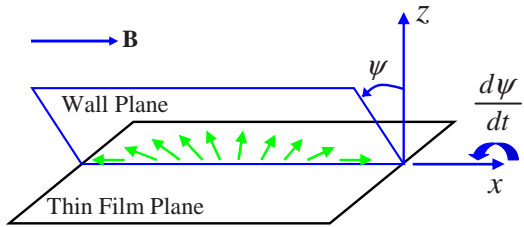}
\caption{\label{fig:TW}(color online). Schematic picture of a
transverse domain wall. The local spin direction changes within a
wall plane which makes an angle $\psi$ with the vertical direction.}
\end{figure}

Next we apply these formulas to the one-dimensional domain wall case
where $\theta$ and $\phi$ only depend on coordinate $x$ of the 2D
plane. We then have $\bm{C}=0$ and $\bm{D}=D\hat{\bm{e}}_x$.  For a
transverse domain wall in the absence of external magnetic field,
the local spins rotate within the 2D plane with the profile
$\theta=\pi/2$ and $\phi(x)=\cos^{-1}\tanh[(x-X)/\lambda]$, where
$X$ denotes the center of the DW and $\lambda$ is the DW width
\cite{malo}. When an external $\textbf{B}$ field is applied, not
only will the DW propagate, but the local spins will also get tilted
out of the x-y plane. As shown in Fig.\ref{fig:TW}, these spins
still lie within a "wall plane" which makes an angle $\psi$ with the
vertical direction. Below Walker's breakdown field, this plane is
fixed, and the angles $\theta$ and $\phi$ are functions of $(x-vt)$,
which makes $\bm{D}=0$. Above Walker's breakdown field, this plane
changes with time (with rate denoted by $d\psi/dt$) \cite{malo},
which leads to a non-zero $\bm D$ field.

If the system is bounded electrically, in a steady state, the DW
motion induced adiabatic force $\hbar(\partial K/\hbar\partial x-D)$
must be balanced by the gradient of electrochemical potential. The
emf along the direction of DW motion, which is the measured voltage
change, is given by
\begin{equation}\label{vxt}
V_x=\frac{\hbar}{e}\int \left(\frac{\partial K}{\hbar\partial
x}-D\right)\mathrm{d}x=\frac{\hbar
}{e}\left(\frac{\mathrm{d}\psi}{\mathrm{d}t}+\frac{2h}{\hbar}\right).
\end{equation}
In the calculation, we use the condition $|h|\ll |J|$, which is
usually the case in experiments on ferromagnetic materials. The
first term on the right hand side represents the so-called \emph{AC
ferro-Josephson effect} proposed by Berger in 1986 based on
phenomenological considerations \cite{berg1}. More careful analyses
on this effect have been done recently using different approaches
\cite{barn,duin,sasl}. The additional term proportional to the field
is due to the difference in Zeeman energy on the two sides of the
DW, which should appear when we suddenly turn on the external field.
However, since the spin relaxation time $\tau_{s}\sim 10^{-12}$s is
much shorter than the characteristic time for DW motion, the voltage
associated with this term cannot be measured in experiment
\cite{pzhang}.

Depending on the thickness and the width of the system, a stable DW
in a thin film can also take a vortex structure \cite{mcmi,klau1}.
In fact, most of the experiments done so far involve vortex walls
rather than transverse walls. In the following, we apply the formula
developed above to study the case of a single vortex in a nanowire.
Assume that the vortex profile is characterized by a core radius
$a$, which is about a few nanometers, and an outer radius $R$, which
is comparable to the wire width $w$. For a vortex centered at
$\bm{X}(t)$, we may approximately write
$\theta=\frac{\pi}{2}\left[1-p\exp(-|\bm{r}-\bm{X}(t)|/a)\right]$
for $|\bm{r}-\bm{X}|<R$ and $\theta=\frac{\pi}{2}$ beyond the outer
radius, $p=\pm 1$ is the polarization. In both cases, we have
$\phi=q\mathrm{arg}(\bm{r}-\bm{X}(t))+c\frac{\pi}{2}$ for
$|\bm{r}-\bm{X}|<R$, where $q=\pm 1$ is the vorticity of the vortex
($q=-1$ is also referred to as antivortex in literature
\cite{naka}), and $c=\pm 1$ indicates its chirality. For a steady
state motion of the vortex, $\theta$ and $\phi$ are functions of
$(\bm{r}-\bm{v}t)$, where $\bm{v}=\dot{\bm{X}}$. Then
\begin{equation}\label{cdrelation}
\bm{D}=-\frac{1}{2}\sin\theta\left[(\bm{v}\cdot\nabla\phi)\nabla\theta-
(\bm{v}\cdot\nabla\theta)\nabla\phi\right]=\bm{C}\times\bm{v},
\end{equation}
which resembles the relation between $\bm{E}$ and $\bm{B}$ fields.
Because $a\ll R\sim w$, these fields are concentrated within the
core region, where
\begin{equation}
\bm{C}=pq\frac{\pi}{4ar}e^{-r/a}\cos\left(\frac{\pi}{2}
e^{-r/a}\right)\hat{\bm{e}}_z,
\end{equation}
and $\bm{D}$ is obtained by Eq.(\ref{cdrelation}). An important
property of the $\bm{C}$ field is that its total flux is a constant,
i.e.,
\begin{equation}\label{cflux}
\int\bm{C}d^2r=\frac{1}{2}\int\sin\theta d\theta
d\varphi\hat{\bm{e}}_z=pq\pi\hat{\bm{e}}_z.
\end{equation}
which is topologically invariant---independent of the detailed
profile of the vortex.

The two-dimensional character of the vortex domain wall makes the
calculation of the induced voltage a bit complicated. Unlike the
one-dimensional transverse wall case, the force field $\bm{D}$ now
has a curl. The gradient of the electrochemical potential can only
cancel the longitudinal part of this force field. Therefore, we need
to solve the Poisson equation $\nabla^2
V=(\hbar/e)\nabla\cdot\bm{D}$ with Neumann boundary condition (no
current leaving the sample). It is seen that the effect of the
$\bm{D}$ field may be regarded as that of an electric dipole
$\textbf{P}$, whose spatial extension is of the size of the core.
The net dipole moment is equivalently given by the integral of the
$\bm D$ field times $\hbar\varepsilon_0/e$. With the help of
Eq.(\ref{cdrelation}) and (\ref{cflux}), $ \textbf{P}
=pq\varepsilon_0\pi(\hbar/e)\hat{\bm{e}}_z\times\bm{v}$, which is
also a topological property of the vortex.

Another complication arises from the Magnus force on a moving vortex
which pushes it in the direction perpendicular to its velocity
\cite{shib}. At low fields, this Magnus force is balanced by the
confining potential of the nanowire such that steady motion is still
along x-direction. In this case, the source term of the Poisson
equation resembles an electric dipole pointing along the
y-direction, so the longitudinal voltage is expected to vanish in
this case (baring nonadiabatic effects).

Above the breakdown field, the confining potential can no longer
balance the Magnus force and the vortex will begin a transverse
motion \cite{klau,he, naka, juny}. When its core hits one edge, the
vortex domain wall transforms into a transverse wall. There is a
range of external field under which this transverse wall propagates
and generates a voltage according to Eq.(\ref{vxt}). But more
probably, another vortex with reversed polarization will be emitted
from the edge, travel across the wire and hit the other edge. These
transformations continue periodically as the DW moves along the wire
\cite{juny}.

\begin{figure}
\includegraphics[width=0.7\columnwidth]{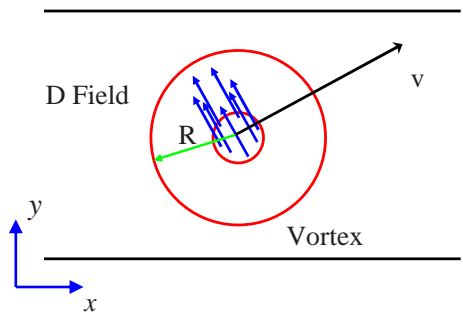}
\caption{\label{fig:VWDfield}(color online). Schematic picture of
the $\bm{D}$ field for a moving vortex. $\bm{D}$ field is
perpendicular to the vortex velocity and concentrated within the
vortex core.}
\end{figure}

When the vortex begins transverse motion, from the relation
$\bm{D}=\bm{C}\times\bm{v}$, we observe that the dipole source gets
rotated to acquire a finite $x$ component. This makes the
longitudinal voltage nonzero (Fig.\ref{fig:VWDfield}). Analytical
expression for the voltage can be obtained within a point-dipole
approximation and using the image charge method.  The longitudinal
voltage drop along the direction of the DW motion is obtained as,
\begin{equation}\label{vx}
V_x=\pi\frac{\hbar}{e}\frac{v_y}{w},
\end{equation}
with $v_y$ being the magnitude of the transverse velocity.

\begin{figure}
\includegraphics[width=0.9\columnwidth]{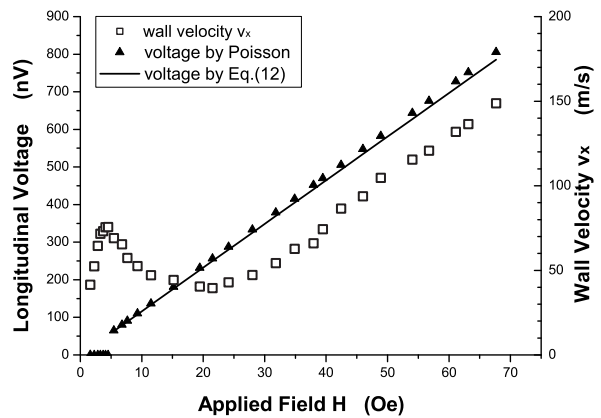}
\caption{\label{fig:Vplot} Numerical results for longitudinal DC
voltage associated with the vortex motion. The triangular data
points are obtained by directly solving the Poisson equation with
$\bar{v}_y=\gamma H w/\pi$ for the sample used in Beach et al.'s
experiment \cite{beac2}. For comparison, a solid line is drawn based
on Eq.(\ref{vxh}). Here also shows the mobility curve measured
experimentally in Ref.\cite{beac2}.}
\end{figure}

This longitudinal voltage drop is proportional to the transverse
speed, and inversely proportional to the wire width. The result is
universal in the sense that it is independent of the detailed wall
profile including its polarization and vorticity. In fact, as
confirmed by our numerical calculations, this result is exact beyond
the point-dipole approximation as long as the core is contained
within the width and is far from the two ends of the sample.
Measurement of this universal result may provide clear
characterization of the vortex motion.

The average value of this voltage depends on the average frequency
of the wall transformation. Recent experiments and simulations
\cite{haya1,juny} suggest that for narrow nanowires, this frequency
is approximately the Larmor frequency. So $\bar{v}_y/w=\gamma
H/\pi$, where $\bar{v}_y$ is time averaged transverse speed,
$\gamma$ is the gyromagnetic ratio and $H$ is the applied magnetic
field strength. In this case,
\begin{equation}\label{vxh}
\bar{V}_x=\frac{\hbar}{e}\gamma H,
\end{equation}
which means the DC part of the voltage signal is universal and only
depends on the applied magnetic field. This result is in good
agreement with the voltage obtained by solving Poisson equation
directly (Fig.\ref{fig:Vplot}). Here we also expect there are small
oscillations around this DC signal with the Larmor frequency.

The motion of the vortex also induces a transverse voltage, which
can also be calculated from the Poisson equation. For a vortex at
the center of the wire, the transverse voltage at the longitudinal
position of the vortex is found to be
\begin{equation}
V_y=\pi pq\frac{\hbar}{e}\frac{v_x}{w}.
\end{equation}
which can be measured through a pair of lateral leads. At low
fields, this voltage is a constant, and one should observe a pulse
of transverse voltage with the above peak value with its width
determined by the wall speed and lead width. Strong oscillations
occur within this pulse above the breakdown field when the vortex or
antivortex executes complicated motion.

So far we have only considered the spin majority carriers and the
sample being disorder free. In real situation, for adiabatic
approximation to be valid, we need $J/(\hbar/\tau)\gg 1$, where
$\tau$ is the carrier mean free time between elastic scattering
events \cite{popp}. This condition ensures the probability of spin
flip due to collision broadening is negligible. For Permalloy thin
films at room temperature, this condition still holds for spin
majority carriers but breaks down for minority carriers. Therefore
the above Berry phase effects disappear for minority carriers
because their phases are randomized due to scattering.

Finally, we give an estimation of the extra voltage due to
nonadiabatic or dissipative effects. Originally obtained from force
balance considerations \cite{berg}, this contribution has been
related to the nonadiabatic spin transfer torque recently
\cite{duin}. This voltage drop can be written as $
V^{na}_x=2M_sR_0\mu_i^{-1}v_x$, where $R_0$ is the ordinary Hall
coefficient and $\mu_i$, called the intrinsic wall mobility, is a
measure of the electrons' contribution to the viscous damping force
on the domain wall. For Permalloy thin films, $M_s=8\times 10^5$A/m,
$R_0=-1.4\times 10^{-10}$m$^3$/C and $\mu_i\simeq 2$m$^2$/C. Then
$V^{na}_x$ is at least 10 times smaller than the adiabatic voltage
above breakdown. However, it is the dominant contribution to the
longitudinal voltage below breakdown.

In summary, we have proposed a general theory for studying electron
dynamics in the presence of moving local spin textures. We find that
the variation of local spin textures gives rise to two topological
fields acting on the conduction electrons as driving forces. Using
this formalism, we reproduce the result for transverse wall motion.
Moreover, universal results are obtained for the voltage induced by
a moving vortex wall and its measurement can be used for detecting
the domain wall motion. Finally, we estimate the nonadiabatic
contributions to the voltage drop which shall be important below
Walker breakdown.

The authors would like to thank Changhai Xu, Weidong Li, Chih-Piao
Chuu, Wang Yao, Dennis P. Clougherty, Shufeng Zhang, Geoffrey S. D.
Beach, Maxim Tsoi and James L. Erskine for valuable discussions. SY
was supported by NSF DMR-0404252, DX was supported by NSF
DMR-0606485, and QN by the Welch Foundation and DOE
(DE-FG03-02ER45958).


\begin{thebibliography}{99}
\bibitem{greg} J. F. Gregg, W. Allen, K. Ounadjela, M.  Viret,
M. Hehn, S. M. Thompson, and J. M. D. Coey, Phys. Rev.  Lett. {\bf
77}, 1580 (1996).
\bibitem{tafu} G. Tatara and H. Fukuyama, Phys. Rev. Lett. {\bf 78}, 3773
(1997).
\bibitem{gork} R. P. van Gorkom, A. Brataas, and G. E. W. Bauer,
  Phys. Rev. Lett. {\bf 83}, 4401 (1999).
\bibitem{duga} V. K. Dugaev, J. Barnas, A. \L{}usakowski, and \L{}. A.
Turski, Phys. Rev. B {\bf 65}, 224419 (2002).

\bibitem{berg} L. Berger, J. Appl. Phys. {\bf 55}, 1954 (1984).
\bibitem{tk} G. Tatara and H. Kohno, Phys. Rev. Lett. {\bf 92}, 086601
(2004).
\bibitem{lizh} Z. Li and S. Zhang, Phys. Rev. B {\bf 70}, 024417
(2004); Phys. Rev. Lett. {\bf 92}, 207203 (2004). S. Zhang and Z.
Li, Phys. Rev. Lett. {\bf 93}, 127204 (2004). Z. Li, J. He, and S.
Zhang, J. Appl. Phys. {\bf 99}, 08Q702 (2006).
\bibitem{thia} A. Thiaville, J. Miltat and J. Vernier, J. Appl.
Phys. {\bf 95} 7049 (2004). A. Thiaville, Y. Nakatani, J. Miltat,
and Y. Suzuki, Europhys. Lett. {\bf 69}, 990 (2005).
\bibitem{barn1} S. E. Barnes and S. Maekawa, Phys. Rev. Lett. {\bf
95}, 107204 (2005).
\bibitem{ohe} J. Ohe and B. Kramer, Phys. Rev. Lett. {\bf 96},
027204 (2006).
\bibitem{bena} M. Benakli, J. Hohlfeld, and A. Rebei,arXiv:0708.2412v1.
\bibitem{yama} A. Yamaguchi, T. Ono, S. Nasu, K. Miyake, K. Mibu, and
T. Shinjo, Phys. Rev. Lett. {\bf 92}, 077205 (2004).
\bibitem{klau}M. Kl\"{a}ui, P.-O. Jubert, R. Allenspach, A. Bischof,
J. A. C. Bland, G. Faini, U. R\"{u}diger, C. A. F. Vaz, L. Vila, and
C. Vouille, Phys. Rev. Lett. {\bf 95}, 026601 (2005).
\bibitem{beac1} G. S. D. Beach, C. Knutson, C. Nistor, M. Tsoi, and
J. L. Erskine, Phys. Rev. Lett. {\bf 97}, 057203 (2006).
\bibitem{haya} M. Hayashi, L. Thomas, Ya. B. Bazaliy, C. Rettner,
R. Moriya, X. Jiang, and S. S. P. Parkin , Phys. Rev. Lett. {\bf
96}, 197207 (2006).

\bibitem{berg1} L. Berger, Phys. Rev. B {\bf 33}, 1572 (1986).
\bibitem{barn} S. E. Barnes, J. Ieda and S. Maekawa, Appl. Phys. Lett. {\bf 89},
122507 (2006). S. E. Barnes and S. Maekawa, Phys. Rev. Lett. {\bf
98}, 246601 (2007).
\bibitem{duin} R. A. Duine, Phys. Rev. B {\bf 77}, 014409 (2008).
\bibitem{sasl} W. M. Saslow, Phys. Rev. B {\bf 76}, 184434 (2007).

\bibitem{berr} M. V. Berry, Proc. R. Soc. London, Ser. A {\bf 392},
45 (1984).
\bibitem{brau} H. B. Braun and D. Loss, Phys. Rev. B {\bf 53}, 3237
(1996).
\bibitem{baza} Y. B. Bazaliy, B. A. Jones, and S. C. Zhang, Phys.
Rev. B {\bf 57}, R3213 (1998).

\bibitem{mcmi} R. D. Mcmichael and M. J. Donahue, IEEE Trans. Magn. {\bf
33}, 4167 (1997).
\bibitem{klau1} M. Kl\"{a}ui, C. A. F. Vaz and J. A. C. Bland, L. J.
Heyderman, F. Nolting, A. Pavlovska, E. Bauer, S. Cherifi, S. Heun,
and A. Locatelli, Appl. Phys. Lett. {\bf 85}, 5637 (2004).

\bibitem{sund} G. Sundaram and Q. Niu, Phys. Rev. B {\bf 59}, 14915
(1999).

\bibitem{malo} A. P. Malozemoff and J. C. Slonczewski, \emph{Magnetic Domain Walls in
Bubble Materials} (Academic Press, New York, 1979).

\bibitem{pzhang} Private communication with S. Zhang.

\bibitem{shib} J. Shibata, Y. Nakatani, G. Tatara, H. Kohno, Y. Otani,
Phys. Rev. B {\bf 73}, 020403(R) (2006).
\bibitem{he} J. He, Z. Li and S. Zhang, Phys. Rev. B {\bf 73}, 184408
(2006).
\bibitem{naka} Y. Nakatani, A. Thiaville and J. Miltat, Nature
Mater. {\bf 2}, 521 (2003).
\bibitem{juny} J.-Y. Lee, K.-S. Lee, S. Choi, K. Y. Guslienko, and
S.-K. Kim, arXiv:cond-mat/07062542. J.-Y. Lee, K.-S. Lee, S. Choi,
K. Y. Guslienko, and S.-K. Kim, Phys. Rev. B {\bf 76}, 184408
(2007).
\bibitem{haya1} M. Hayashi, L. Thomas, C. Rettner, R. Moriya and S.
S. P. Parkin, Nature Physics {\bf 3}, 21 (2007).

\bibitem{beac2} G. S. D. Beach, C. Nistor, C. Knutson, M. Tsoi, and
J. L. Erskine, Nat. Mater. {\bf 4}, 741 (2005).

\bibitem{popp} M. Popp, D. Frustaglia, and K. Richter, Phys. Rev. B {\bf
68}, 041303(R) (2003).

\end{thebibliography}
\end{document}